\RequirePackage{fix-cm}
\documentclass[conference]{IEEEtran}
\usepackage[latin9]{inputenc}
\usepackage{amsmath}\usepackage{amssymb}\usepackage{graphicx}
\usepackage{tikz}\usetikzlibrary{automata,arrows,positioning,calc}

\makeatletter

\usepackage{epsfig}\usepackage{amsfonts}\usepackage{array}\usepackage{dcolumn}\usepackage{psfrag}\usepackage{amsfonts}\usepackage{accents}\usepackage{booktabs}
\usepackage{cite}\usepackage{subfigure}\usepackage[all]{xy}
\usepackage{dsfont}\usepackage{url}\usepackage{xcolor}\usepackage[nolist]{acronym}
\usepackage[footnote,draft,silent,nomargin]{fixme}
\usepackage{algorithm}
\usepackage[noend]{algpseudocode}
\usepackage{amsthm}
\usepackage{mathtools}
\usepackage{xfrac}
\usepackage{booktabs}
\graphicspath{{./figures/}}

\acrodef{SINR}{Signal to Interference and Noise Ratio}
\acrodef{CDF}{Cumulative Distribution Function}
\acrodef{PDF}{Probability Density Function}
\acrodef{MTC}{Machine-type Communications}
\acrodef{M2M}{Machine-to-Machine}
\acrodef{CTMC}{Continuous-Time Markov Chain}
\acrodef{RV}{Random Variable}
\acrodef{MTTR}{Mean Time to Restoration}
\acrodef{TTI}{Transmission Time Interval}

\DeclareMathOperator*{\argmin}{arg\,min}
\renewcommand{\vec}[1]{\underline{#1}}

\makeatother

\begin{document}

\title{Latency Analysis of Systems with Multiple Interfaces for Ultra-Reliable M2M Communication}

\author{
Jimmy~J.~Nielsen and Petar~Popovski\\
APNET section, Department of Electronic Systems, Aalborg University, Denmark\\
\{jjn,petarp\}@es.aau.dk\\%
}

\maketitle
\begin{abstract} 
One of the ways to satisfy the requirements of ultra-reliable low latency communication for mission critical \acf{MTC} applications is to integrate multiple communication interfaces. In order to estimate the performance in terms of latency and reliability of such an integrated communication system, we propose an analysis framework that combines traditional reliability models with technology-specific latency probability distributions.
In our proposed model we demonstrate how failure correlation between technologies can be taken into account. We show for the considered scenario with fiber and different cellular technologies how up to 5-nines reliability can be achieved and how packet splitting can be used to reduce latency substantially while keeping 4-nines reliability. The model has been validated through simulation. 
\end{abstract}

\section{Introduction}
Some of the defining features of the upcoming 5G networks are native support for \ac{M2M}, ultra-high reliability, and integration of different communication technologies \cite{andrews2014will} \cite{monserrat2015metis}. Specifically, 5G is expected to cover new use cases such as mission critical \ac{MTC}, whose requirements exceed the capabilities of current technologies. Reliability requirements in terms of packet loss rates may be as high as 5-nines to 9-nines, while also the acceptable latency may be down to few milliseconds \cite{ratasuk2015recent}. While there are proposals for how to decrease the end-to-end latency in cellular systems, e.g., by reducing the \ac{TTI} \cite{lahetkangas2014achieving,tullberg2014towards}, the very high levels of reliability can most feasibly be achieved by the integration of multiple communication technologies \cite{dahlman20145g}.

The use of multiple communication technologies is conceptually very similar to many existing multipath protocols that can increase reliability of communication between two end hosts \cite{qadir2015exploiting}. However, given the strict latency requirements of mission critical MTC, we cannot use protocols that rely on error detection and retransmission.  Instead we focus on path diversity \cite{apostolopoulos2000reliable}, with the additional constraint that each path must use a different communication interface. While this clearly improves reliability, the bandwidth usage increases. If the transmitted information is split in parts and only some parts are sent over each interface, it is possible to trade-off bandwidth usage and reliability according to the targeted application.
For this, it is necessary to estimate the reliability of the multi-interface communication system, while accounting for component failures as well as the achievable latency of the different interfaces that may use different technology.

In this paper we propose an analysis framework that allows us to estimate the latency and reliability performance of different multi-path transmission configurations by combining traditional reliability engineering methods with technology specific latency-reliability characteristics.

First, in sec.~\ref{sec:reliability_in_comms} we introduce the concepts of reliability in communications, which is followed by the system model for the considered mission critical MTC scenario in sec.~\ref{sec:system_model}. Hereafter, we present the proposed analysis and modeling framework in sec.~\ref{sec:reliability_miftx}, which is evaluated numerically in sec.~\ref{sec:results}. Conclusion and outlook are given in sec.~\ref{sec:conclusion}.

\section{Reliability of Communications}\label{sec:reliability_in_comms}
In reliability engineering, reliability is the probability that a system can provide uninterrupted delivery of acceptable service for a given mission time \cite{rausand2004system}. 
The most basic approach to estimating reliability is to assume exponential failure times and then to use parallel and series systems theory or a \ac{CTMC} state model to calculate the probability of no system failures during the mission time \cite{rausand2004system}.
These methods alone do, however, not let us account for the typical latency and jitter of communication technologies. In communications, reliability is typically defined as a system's ability to deliver some amount of information (data packet) within a certain (application dependent) deadline \cite{strom20155g}.

\begin{figure}[bt]
	\centering
	\includegraphics[width=\linewidth]{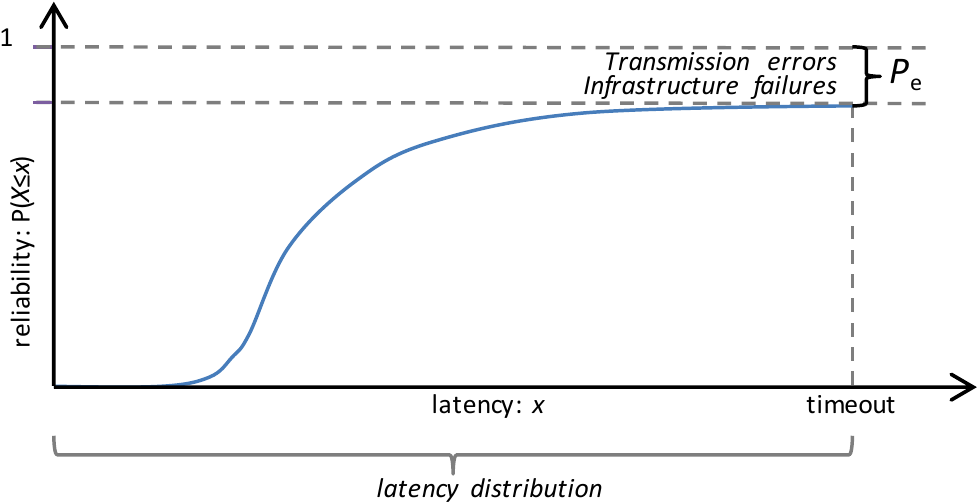}
	\caption{Conceptual illustration of latency-reliability function.}
	\label{fig:latency_cdf_conceptual}
\end{figure}

For representing the latency characteristics of an interface, we use the latency \ac{CDF} \cite{strom20155g}. For the example in Fig. \ref{fig:latency_cdf_conceptual}, the blue curve is the probability that a packet is delivered from source to destination within a certain latency deadline. That is, for a given latency value $x$ on the x-axis, the reliability, i.e., the probability that the packet latency $X \leq x$ can be read off the y-axis.
Such a curve can be produced from network monitoring measurements, e.g., by continually pinging the remote host and recording success rate and latency.

The shape of the curve depends on two factors: 1) the variability of latency, i.e., the \emph{latency distribution}, due to factors such as medium access, routing, queuing and processing delays, and 2) the loss probability due to various failures between two end hosts. Specifically, this loss can be caused by \emph{transmission errors} such as low \ac{SINR}, network access overload, packet drops, and congestion, or by \emph{Infrastructure failures} such as cable fractures, equipment malfunction, or power outage.
The combined loss probability resulting from transmission errors and infrastructure failures is denoted $P_\text{e}$ in the figure. Unless $P_\text{e}=0$, the latency distribution curve never reaches 1, meaning that the curve is technically not a \ac{CDF}. We will refer to it as \textit{latency-reliability function} in the rest of this paper.

\section{System model}
\label{sec:system_model}
\begin{figure}[htb]
	\centering
	\includegraphics[width=\linewidth]{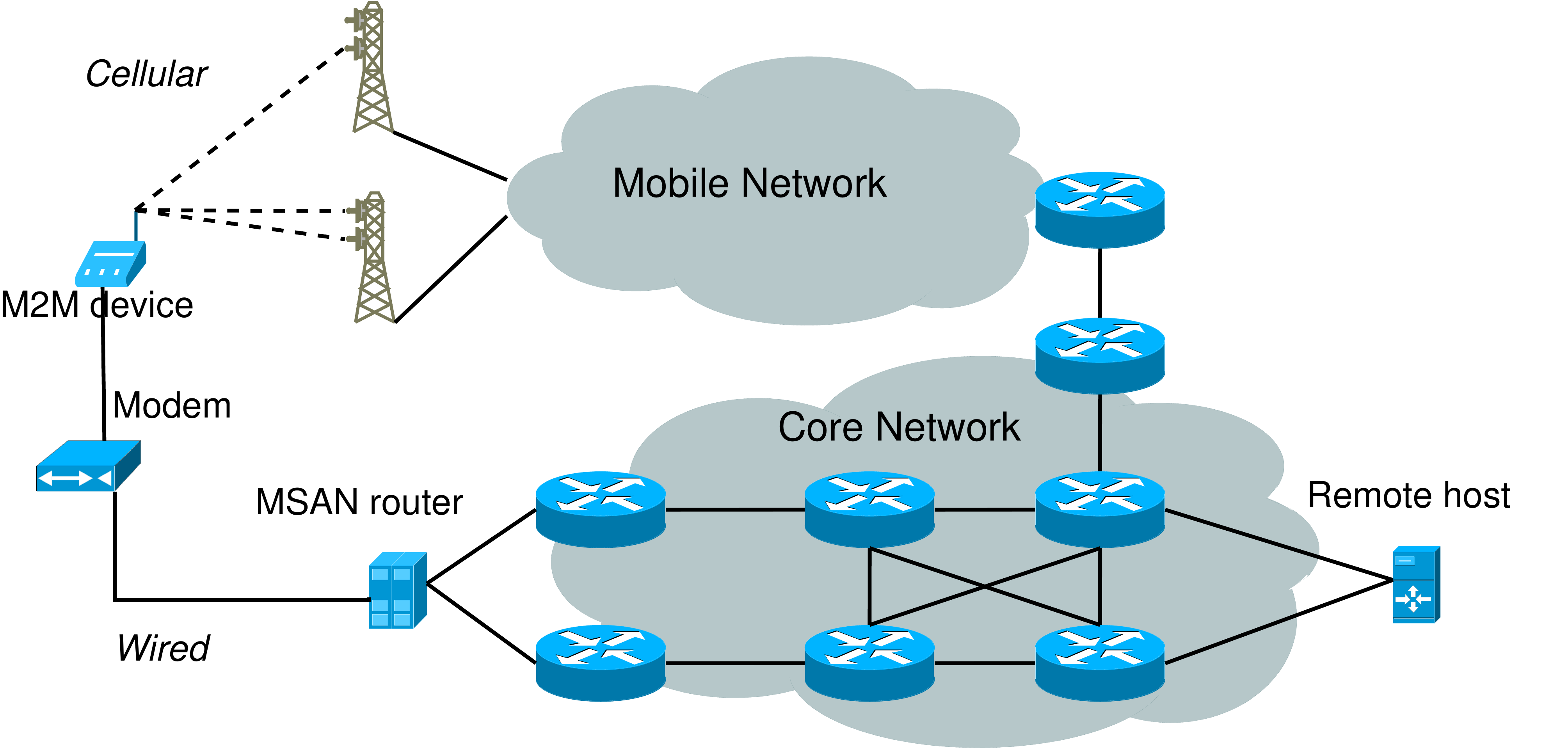}
	\caption{Multiple paths between M2M device (left) and remote host (right).}
	\label{fig:network_diagram}
\end{figure}

We consider an M2M device that needs to communicate reliably with a specific end-host, e.g., a monitoring device reporting measurements, status and alarm messages to a control unit.
The M2M device has $N$ communication interfaces (wired and cellular) available to reach the end-host, as pictured in Fig. \ref{fig:network_diagram}. For each interface, a latency reliability function as in Fig. \ref{fig:latency_cdf_conceptual} is available. Notice that some technologies are mostly independent while cellular connections that share the same base station may have a higher degree of failure correlation.

\subsection{Transmission Strategies}
When transmitting information from the source to the remote host, one or more of the available interfaces can be used simultaneously.
In addition to packet cloning, where a full copy of the payload is sent on each interface, we consider different splitting approaches in which the source data packet is split into $n$ fragments and subsets of these fragments are then transmitted over each interface.
The possibility for reducing the bandwidth usage is useful with cellular technologies, since 1) resources in licensed bands are scarce and costly, and 2) the reduction of payload size reduces latency.
Due to the dependence on the packet size, it is necessary to characterize the relationship between payload size and latency distribution. 
We specify the latency-reliability function of interface $i$ as $F_i(x,B)$. This gives the probability of being able to transmit a data packet of $B$ bytes from a source to a destination via interface $i$ within a latency deadline $x$. 
In other words, the value of $F_i(x,B)$ is the achievable reliability ($P(X \leq x$, c.f., Fig. \ref{fig:latency_cdf_conceptual}) for a latency value $x$ and payload size $B$.
Further, by $P_\text{e}^{(i)}$ we refer to $P_\text{e}$ (defined in Fig. \ref{fig:latency_cdf_conceptual}) for the $i$th interface.

\section{Reliability of Multi-Interface Transmissions}\label{sec:reliability_miftx}
This section presents the proposed methods for evaluating the reliability of multi-interface transmissions. 

\subsection{Parallel transmissions over independent interfaces}
For transmissions using packet cloning over $k$ interfaces that can justifiably be considered independent, e.g., an optical fiber link and a cellular link\footnote{While the telco core network may be shared for such two options, the core network has several redundancy mechanisms to ensure very high levels of reliability, and we will therefore not focus on its impact in the following.}, we use the traditional parallel systems \cite{rausand2004system} method to combine the latency-reliability functions:
\begin{equation}
	F_\text{$k$-par}(x,B) = 1-\prod\limits_{i=1}^k (1-F_i(x,\kappa_i B))\label{eq:f_k_par}.
\end{equation}
However, for cellular networks whose base station equipment may be located in the same tower, complete independence cannot be assumed, in case of common root cause failures.

\subsection{Parallel transmissions with failure correlation}
Correlated failures can be modeled for example by using a \ac{CTMC} state model. For such a model, we calculate the combined latency-reliability function as:
\begin{equation}
		F_{k\text{-dep}}(x,B) = \sum\limits_{s=1}^{L} \pi_s \cdot F_s^\text{st}(x,B), \label{eq:F_dep}
\end{equation}
where $L$ is the number of states in the \ac{CTMC}, $\pi_s$ is the steady-state probability of state $s$ in the \ac{CTMC}, and $F_s^\text{st}(x,B)$ characterizes the latency-reliability function of state $s$. Notice the use of superscript ''st'' to not confuse the latency-relibility function of a state ($F_s^\text{st}$) with a latency-reliability function of an interface ($F_i$).
%
Since the latency-reliability function to use for a given system state ($F_s^\text{st}(x,B)$) depends on the considered system model and whether packet cloning or splitting is used, we need to first introduce the specific case study that we consider in this paper before we can define it further.

\section{Case study: A system with three interfaces}
\begin{figure}[htb]
	\centering
	\resizebox{\linewidth}{!}{
\begin{tikzpicture}[->, >=stealth', auto, semithick, node distance=3.5cm]
\tikzstyle{every state}=[fill=white,draw=black,thick,text=black,scale=0.9,align=center]
\node[state]    (A)[fill=green!20]   {\textbf{1}\\All OK};
\node[state]    (C)[right of=A,fill=yellow!20]   {\textbf{3}\\C2\\fail};
\node[state]    (B)[above of=C,fill=yellow!20]   {\textbf{2}\\C1\\fail};
\node[state]    (D)[below of=C,fill=yellow!20]   {\textbf{4}\\Fi\\fail};
\node[state]    (E)[below of=D,fill=orange!20]   {\textbf{5}\\BS\\fail};

\node[state]    (G)[right of=B,fill=orange!20]   {\textbf{7}\\C1+Fi\\fail};
\node[state]    (F)[above of=G,fill=orange!20]   {\textbf{6}\\C1+C2\\fail};
\node[state]    (H)[below of=G,fill=orange!20]   {\textbf{8}\\C1+BS\\fail};
\node[state]    (I)[below of=H,fill=orange!20]   {\textbf{9}\\C2+Fi\\fail};
\node[state]    (J)[below of=I,fill=orange!20]   {\textbf{10}\\C2+BS\\fail};
\node[state]    (K)[below of=J,fill=red!30]      {\textbf{11}\\Fi+BS\\fail};

\node[state]    (L)[right of=G,fill=red!30]   {\textbf{12}\\C1+C2+Fi\\fail};
\node[state]    (M)[below of=L,fill=orange!20]{\textbf{13}\\C1+C2+BS\\fail};
\node[state]    (N)[below of=M,fill=red!30]   {\textbf{14}\\C1+Fi+BS\\fail};
\node[state]    (O)[below of=N,fill=red!30]   {\textbf{15}\\C2+Fi+BS\\fail};
\path[<->]
(A) edge[bend left=10]     node{C1}         (B)
(A) edge[bend left=0]     node{C2}         (C)
(A) edge[bend left=0]     node{Fi}         (D)
(A) edge[bend right=10]     node{BS}         (E)

(B) edge[bend left=10]     node[pos=0.1]{C2}         (F)
(B) edge[bend left=0]     node[pos=0.2]{Fi}         (G)
(B) edge[bend left=0]     node[pos=0.08]{BS}         (H)

(C) edge[bend right=10]     node[pos=0.1]{C1}         (F)
(C) edge[bend left=10]     node[pos=0.1]{Fi}         (I)
(C) edge[bend left=10,below]     node[pos=0.05]{BS}         (J)
(D) edge[bend right=10]     node[pos=0.1]{C1}         (G)
(D) edge[bend left=0]     node[pos=0.1]{C2}         (I)
(D) edge[bend left=5]     node[pos=0.1]{BS}         (K)
(E) edge[bend right=10]     node[pos=0.05]{C1}         (H)
(E) edge[bend left=0]     node[pos=0.11]{C2}         (J)
(E) edge[bend right=10]     node[pos=0.08]{Fi}         (K)

(F) edge[bend left=20]     node[pos=0.08]{Fi}         (L)
(F) edge[bend left=0]     node[pos=0.08]{BS}         (M)

(G) edge[bend left=0]     node[pos=0.2]{C2}         (L)
(G) edge[bend left=0]     node[pos=0.08]{BS}         (N)

(H) edge[bend left=0]     node[pos=0.2]{C2}         (M)
(H) edge[bend left=0]     node[pos=0.1]{Fi}         (N)

(I) edge[bend left=0]     node[pos=0.06]{C1}         (L)
(I) edge[bend right=10]     node[pos=0.1]{BS}         (O)

(J) edge[bend left=0]     node[pos=0.06]{C1}         (M)
(J) edge[bend left=0]     node[pos=0.2]{Fi}         (O)

(K) edge[bend left=0]     node[pos=0.05]{C1}         (N)
(K) edge[bend right=10,below]     node[pos=0.18]{C2}         (O)

;
\end{tikzpicture}}
	\caption{CTMC model of states in the three interface system. Colors indicate the number of interfaces \textit{up/down} as: Green: 3/0, yellow: 2/1, orange: 1/2, red: 0/3. An arrow represents a failure rate in the right direction and restoration rate in the left direction, e.g., $\lambda_{C1}$ and $\mu_{C1}$ between states 1 and 2.}
	\label{fig:15-state_diagram}
\end{figure}
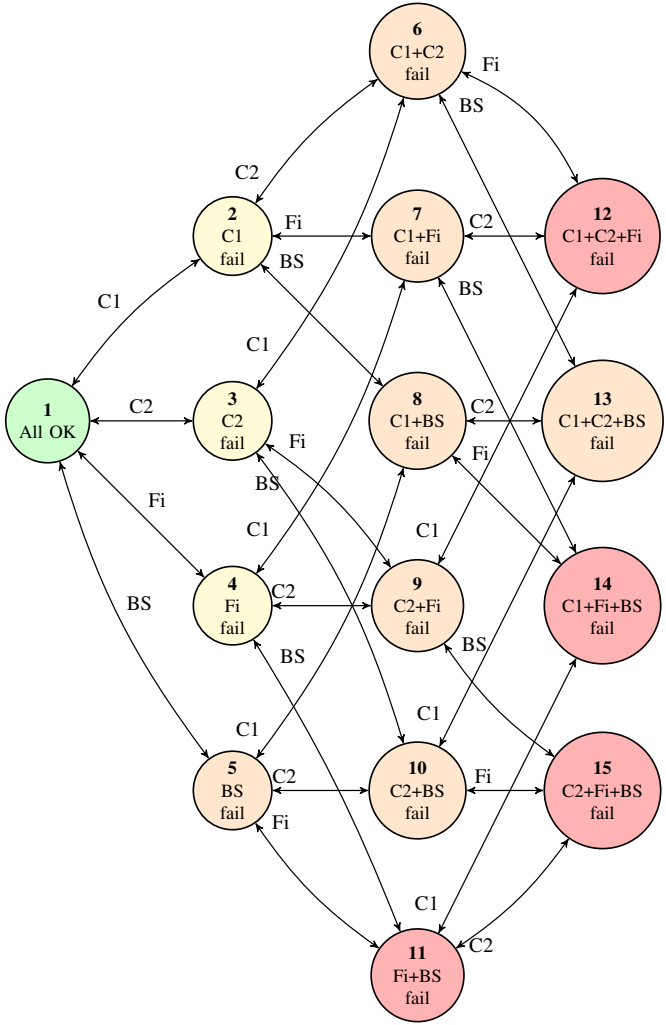

We assume that the M2M device in Fig. \ref{fig:network_diagram} is connected by fiber and has two cellular interfaces, denoted by $C1$ and $C2$. This is a typical mission critical MTC use case from smart grid systems \cite{stefanovic2014sunseed}. The CTMC in Fig. \ref{fig:15-state_diagram} shows the different modes of operation considered for the case study. 
For each of the considered transmission strategies, we present a short description and define the state-specific latency-reliability functions $F_s^\text{st}(x,B)$ for $s={1,2,\ldots, L}$ that are represented by the vector $\vec{F}^\text{st}(x,B)$. 

In the following, the shorthand notation $\hat{F}_i = F_i(x,\kappa_i B)/A_i$ is the latency reliability function normalized by the availability $A_i$, thereby making $\hat{F}_i$ a CDF. We do this because we have used $A_i=1-P_\text{e}^{(i)}$ in the parametrization of the \ac{CTMC} model.
By normalizing $P_\text{e}$ out of the latency-reliability function and including it in the CTMC we enable the use of probability theory for the following analysis.
The value of $\kappa_i$ depends on the transmission strategy used, as specified in Table \ref{tab:split_strategies}.
Further, for compact notation of interface-specific latency-reliability functions, we 
let $i\!=\!1$ represent fiber, $i\!=\!2$ is $C1$, and $i\!=\!3$ is $C2$. Illustrations of the strategies and packet size scaling parameters $\kappa_i$, are shown in Table \ref{tab:split_strategies} and they are explained in the next section. Notice that the \ac{CTMC} system model in Fig. \ref{fig:15-state_diagram} is used with all three strategies.

\begin{table}[htb]
\centering
\caption{Packet splitting strategies and parameters}
\setlength{\tabcolsep}{3pt}
\begin{tabular}{lcccccccccc}
\toprule
\multicolumn{1}{l}{}                 & \multicolumn{3}{c}{\textbf{cloning}}             && \multicolumn{3}{c}{\textbf{2-of-3}}             &                         & \multicolumn{2}{c}{\textbf{weighted}}  \\ \cmidrule{2-4} \cmidrule{6-8} \cmidrule{10-11}
\multicolumn{1}{l}{} 				& \multicolumn{1}{c}{$B_1$} & \multicolumn{1}{c}{$B_2$} & \multicolumn{1}{c}{$B_3$} & & \multicolumn{1}{c}{$B_1$} & \multicolumn{1}{c}{$B_2$} & \multicolumn{1}{c}{$B_3$} & & \multicolumn{2}{c}{--} \\ \midrule
\multicolumn{1}{l}{\textit{Fiber (1)}}           & x                      & x                      & \multicolumn{1}{c}{x} & & x                      & x                      &   & & \multicolumn{2}{c}{--} \\ 
\multicolumn{1}{l}{\textit{C1 (2)}}              & x                      & x                      & \multicolumn{1}{c}{x} & &                        & x                      & x & & \multicolumn{2}{c}{--} \\ 
\multicolumn{1}{l}{\textit{C2 (3)}}              & x                      & x                      & \multicolumn{1}{c}{x} & & x                      &                        & x & & \multicolumn{2}{c}{--} \\ \midrule
$\kappa_\text{1}$ & \multicolumn{3}{c}{$1$} & & \multicolumn{3}{c}{$\sfrac{2}{3}$} & & \multicolumn{2}{c}{$1$} \\
$\kappa_{2}$ & \multicolumn{3}{c}{$1$} & & \multicolumn{3}{c}{$\sfrac{2}{3}$} & & \multicolumn{2}{c}{$\gamma$} \\
$\kappa_{3}$ & \multicolumn{3}{c}{$1$} & & \multicolumn{3}{c}{$\sfrac{2}{3}$} & & \multicolumn{2}{c}{$1\!-\!\gamma$} \\
\bottomrule
\end{tabular}
\parbox[t]{.65\linewidth}{
\vspace{.5\baselineskip}
\textit{\footnotesize{Crosses indicate on which interfaces packet fragments are sent. $B_j$ is the $j$th third of the payload.}}
}
\label{tab:split_strategies}
\end{table}

\subsection{Packet cloning on three interfaces}
For each state in Fig. \ref{fig:15-state_diagram}, we need to specify how the interfaces' latency-reliability functions $F_i(x,B)$ should be combined. In states where more than one interface is available, the latency is given by the first arriving packet. Let the independent \acp{RV} $X_1, ..., X_k$ represent the latency of each of the $k\in\{1,2,3\}$ interfaces. The latency CDF of the first arriving is known to be $F_\text{min}=1-\Pi_{j=1}^k(1-F_{j})$. Thus, the $F_i(x,B)$ functions are combined as shown in Table \ref{tab:P_vectors}.
The resulting latency-reliability function is computed using \eqref{eq:F_dep}.

\begin{table*}[t]
\centering
\caption{Latency-reliability function vector $\vec{F}^\mathrm{st}(x,B)$ for the considered strategies, with $\hat{F}_i = {F_i(x,\kappa_i B)}/{A_i}$.}
\begin{tabular}{ccccc}
\toprule
\textbf{cloning} & & \textbf{2-of-3} & & \textbf{weighted} \\ \cmidrule{1-1} \cmidrule{3-3} \cmidrule{5-5}
$\begin{bmatrix}
     1-(1-\hat{F}_{1})(1-\hat{F}_{2})(1-\hat{F}_\text{3}) \\
     1-(1-\hat{F}_{2})(1-\hat{F}_\text{1}) \\
     1-(1-\hat{F}_{2})(1-\hat{F}_\text{1}) \\
     1-(1-\hat{F}_{3})(1-\hat{F}_{2}) \\
     \hat{F}_\text{1} \\
     \hat{F}_\text{1} \\
     \hat{F}_{3} \\
     \hat{F}_\text{1} \\
     \hat{F}_{2} \\
     \hat{F}_\text{1} \\
     0 \\
     0 \\
	 \hat{F}_\text{1} \\
     0 \\
     0
    \end{bmatrix}$
& &
$\begin{bmatrix}
	 F_\text{A} + F_\text{B} + F_\text{C} + F_\text{D}\\
     \hat{F}_\text{1}\hat{F}_{3} \\
     \hat{F}_\text{1}\hat{F}_{2} \\
     \hat{F}_{2}\hat{F}_{3} \\
     0 \\
     0 \\
     0 \\
     0 \\
     0 \\
     0 \\
     0 \\
     0 \\
     0 \\
     0 \\
     0
    \end{bmatrix}$
& &
$\begin{bmatrix}
      1\!-\!(1-\hat{F}_{1})(1-(\hat{F}_{2}\hat{F}_{3}) \\
     \hat{F}_{1} \\
     \hat{F}_{1} \\
     \hat{F}_{2}\hat{F}_{3} \\
     \hat{F}_{1} \\
     \hat{F}_{1} \\
     0\\
     \hat{F}_{1} \\
     0\\
     \hat{F}_{1} \\
     0\\
     0\\
     \hat{F}_{1} \\
     0\\
     0
    \end{bmatrix}$ \\
    \bottomrule
\end{tabular}
\label{tab:P_vectors}
\end{table*}

\subsection{2-of-3 packet splitting on three interfaces}
This strategy is based on the three-interface transmissions above, however, instead of transmitting a full replica of the source message on each interface, a 2-of-3 splitting is used so that each interface carries a fragment that contains only $\sfrac{2}{3}$ of the information of the source message. The fragments sent on each interface are composed so that the reception of two different fragments allows the source message to be successfully decoded, as sketched in Table \ref{tab:split_strategies}.
Consequently, the state-specific latency-reliability functions are different than for packet cloning. 
In state 1, to compute the probability of receiving at least 2 fragments within a latency value $x$, we need to consider the four ways in which this can happen. Either all three transmitted fragments are received before $x$ or any two of the three fragments are received before $x$.
The CDFs of these four cases, arbitrarily named A--D, are:
\begin{equation}
	\begin{matrix*}[l]
		F_\text{A} = \hat{F}_1\hat{F}_2\hat{F}_3 &\quad&  F_\text{B} = \hat{F}_1\hat{F}_2(1-\hat{F}_3)\\
		F_\text{C} = \hat{F}_1(1-\hat{F}_2)\hat{F}_3 &\quad& F_\text{D} = (1-\hat{F}_1)\hat{F}_2\hat{F}_3
	\end{matrix*}
\end{equation}
For the CDF of state 1 we use their sum as shown in Table \ref{tab:P_vectors}. On a side note, notice that if we have identical interfaces such that $F_x=\hat{F}_1=\hat{F}_2=\hat{F}_3$ then the expression for the CDF of state 1 simplifies to:
$$3 F_x^2 (1-F_x) + F_x^3,$$
which equals the formula for reliability of a 2-out-of-3 system \cite{rausand2004system}.
For states $2-4$, we use that the second fragment is the last and that its latency CDF is $F_\text{max}=\Pi_{j=1}^k(F_j)$ \cite{ross1996stochastic}.

\begin{table}[bt]
	\centering
	\caption{Case study failure and restoration rates}
	\begin{tabular}{lccc}
	\toprule
		 				& A (availability) & $\lambda$ (f/week) & $\mu$ (r/week) \\ \cmidrule{2-4}
		Cellular (C1, C2)	& 0.98 	& 1.0013	 & 50.4 (200 min/r) \\
		Fiber (fi) 			& 0.998 & 0.0561	 & 28 (6 hrs/r) \\
		Base station (BS) 	& 0.9995 & 0.0267 	 & 50.4 (200 min/r) \\ \bottomrule
	\end{tabular}
	\label{tab:f_and_r_rates}
\end{table}

\begin{table}[bt]
	\centering
	\caption{Linear regression parameters from RTT measurements}
	\label{tab:pl_to_rss_regression}
	\begin{tabular}{lccccc}
	\toprule
	 			 & GPRS & EDGE & UMTS & HSDPA & LTE \\ \cmidrule{2-6}
	 	$\alpha$ & 0.70 & 0.46 & 0.43 & 0.35 & 0.0067 \\
		$\beta$  & 400 & 230 & 200 & 178 & 41 \\ \bottomrule
	\end{tabular}
\end{table}

\subsection{Weighted packet splitting on three interfaces}
This approach is similar to the above except that the distribution of fragments is different. We acknowledge that fiber will likely not be bandwidth limited as a cellular interface, thus, we can send a full copy via fiber, and second we divide the load on the cellular networks more fairly, according to their latency characteristics. Specifically, we want to find the optimal splitting threshold $\gamma$ that minimizes the expected latency in the cellular interfaces, i.e., we want to solve the optimization problem:
\begin{equation}
\min_\gamma \left(\mathbb{E}[\max(X_2,X_3)]\right),
\end{equation}
where $X_2$ and $X_3$ are \acp{RV} representing the latencies of the two cellular interfaces. As described in Table \ref{tab:split_strategies}, we transmit a packet of $\gamma B$ bytes on interface 2 and a packet of $(1-\gamma) B$ bytes on interface 3. Assuming that $X_2$ and $X_3$ are independent and that their \acp{CDF} are $F_{X_2}$ and $F_{X_3}$, respectively, we can characterize the CDF of their maximum as:
\begin{equation}
	F_\text{max}(x,B,\gamma) = F_2(x,\gamma B) F_3(x,(1-\gamma) B).
\end{equation} 
By differentiation of $F_\text{max}$ we get the corresponding \ac{PDF} $f_\text{max}$, and from this we can write out the expectation of the \acp{RV} as:
\begin{equation}\label{eq:E_max}
	\mathbb{E}[\max(X_2,X_3)] = \int\limits_0^\infty x f_\text{max}(x,B,\gamma) \mathrm{d}x
\end{equation}
Thus, our optimization problem is:
\begin{equation}\label{eq:gamma_opt}
	\gamma_\text{opt} = \argmin\limits_{\gamma \in [0,1]} \Big( \int\limits_0^\infty x f_\text{max}(x,B,\gamma)\mathrm{d}x\Big).
\end{equation}

In the next section we use this expression to numerically determine the optimal value of $\gamma$ for the evaluation scenarios.
The latency-reliability function vector $\vec{F}^\mathrm{st}(x,B)$ for the \textit{weighted} strategy is shown in Table \ref{tab:P_vectors} and here we use that the CDF of the latency of the last arriving is $F_\text{max}=\Pi_{j=1}^k(F_{j})$.

\section{Evaluation Scenario}
For the numerical results we will consider a specific scenario with fiber and cellular technologies C1 and C2 being HSDPA and EDGE, respectively.
The considered technologies are using the reliability specifications shown in Table \ref{tab:f_and_r_rates}.
Notice that C1 and C2 can fail simultaneously either due to a common BS failure or independently.


\begin{figure*}[tb]
	\centering
	\includegraphics[width=\linewidth]{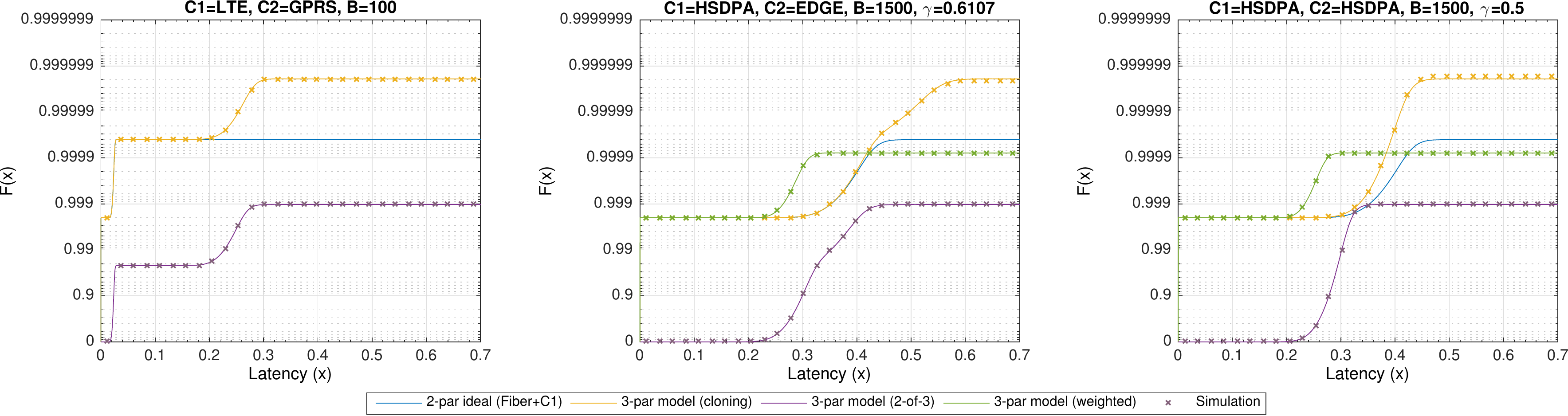}
	\caption{Latency-reliability curves for the considered transmission strategies with different interface configurations and different packet sizes.}
	\label{fig:cdfs_modes}
\end{figure*}

While it is a well-known that the distribution of latency measurements is usually long-tailed \cite{borella1997self}, we assume for simplicity that the latency distribution follows a Gaussian distribution. Specifically, we assume that the latency of transmissions of packet size $B$ through a specific interface/path Gaussian distributed with mean $\mu$ defined as:
\begin{equation}\label{eq:latency_B}
	\mu = \frac{\alpha \cdot B + \beta}{2} [ms]
\end{equation}
and due to lack of information about the distribution, we assume $\sigma = \frac{\mu}{10}~[ms]$.
The parameters $\alpha$ and $\beta$ characterize the assumed linear relationship between packet size and delay for an interface. The values of $\alpha$ and $\beta$ are shown in Table \ref{tab:pl_to_rss_regression}. The values are derived from field measurements conducted by Telekom Slovenije within the SUNSEED project \cite{sunseed2014web}.

For each scenario, the optimal value of $\gamma$ as defined in \eqref{eq:gamma_opt} was found numerically for the one-dimensional convex optimization problem.


For evaluating the resulting performance of the considered transmission modes, actual data on \ac{MTTR} and availability levels of different technologies has been used. From these numbers, the unspecified failure and restoration rates have been determined. The approach to parametrize the CTMC model is explained in the Appendix.
Table \ref{tab:f_and_r_rates} presents the failure and restoration rates used in the numerical evaluation.

\section{Numerical results}
\label{sec:results} 
With failure and restoration rates fully specified, the resulting latency-reliability performance is calculated using the methods outlined in sec.~\ref{sec:reliability_miftx}. The results denoted \textit{2-par, ideal} assume independence between interfaces, whereas the ones denoted \textit{3-par} are using the \ac{CTMC} model. The different model results have been verified using Matlab-based simulation. We first simulated the transitions between states in the \ac{CTMC} model in Fig. \ref{fig:15-state_diagram} with exponential sojourn times given from the rates in Table \ref{tab:f_and_r_rates}.
Hereafter we replayed the state sequence and for every 1 min simulation time, a random Gaussian latency value was drawn for the interfaces available in the current state.  Depending on the required packet fragments of the strategy either a transmission latency or timeout value resulted. The CDF of these values is shown in Fig. \ref{fig:cdfs_modes}.

In all plots in Fig. \ref{fig:cdfs_modes} we see that the \textit{cloning} strategy, which uses a total bandwidth of $3B$, achieves the highest reliability of more than 5-nines for higher latency values. On the contrary, the \textit{2-of-3} strategy that uses a total bandwidth of $3\cdot \frac{2}{3}B=2B$ has the worst reliability of all strategies, since it cannot work with fiber alone, but needs at least one working cellular link as well.
In the leftmost plot with LTE and GPRS for $B=100$ bytes, the weighted strategy is not applicable, since LTE has a 10 times lower latency than GPRS and splitting the payload across the two interfaces therefore does not make sense\footnote{A payload size of more than $B\!>\!{(\beta_\text{LTE}\!-\!\beta_\text{GPRS})}/{\alpha_\text{LTE}}\!\approx\!53600$ bytes is needed before assistance from GPRS is meaningful.}. 
In the center and rightmost plots where $B=1500$ bytes and the used HSDPA and EDGE have more similar latency, there is a latency reduction of 25-30\% at the 4-nines level compared to cloning on 2 and 3 interfaces (blue and yellow curves). For bigger packets even larger reductions can be expected. Keeping in mind that the \textit{weighted} strategy uses a total bandwidth of $(1\!+\!\gamma\!+\!(1\!-\!\gamma))B=2B$ like \textit{2-of-3}, makes it the better choice.
Generally, it is interesting that traditional series, parallel, and $k$-out-of-$n$ reliability models are applicable and very accurate for modeling reliability of packet splitting strategies.

\section{Conclusions and Outlook}
\label{sec:conclusion} 
It is expected that 5G technologies must integrate various communication technologies to provide ultra-high reliability. To estimate the latency and reliability of such an integrated communication system, we propose an analysis framework that combines traditional reliability models with technology-specific latency probability distributions.
Any technologies can be used with the proposed framework given latency curves of interfaces and a model of system states. More advanced models than the used \acl{CTMC} can be used, for example resulting from fault tree analysis. 

In a case study based on availability and latency measurements from the field, we compared different transmission strategies and showed how weighted packet splitting over the cellular interfaces can give substantial latency reductions while maintaining 4-nines reliability. In comparison, using packet cloning on three interfaces yielded 5-nines reliability for higher latency values. Finally, we showed that traditional series, parallel, and $k$-out-of-$n$ reliability models can accurately model the reliability of packet splitting strategies.

\section*{Acknowledgment}
This work is partially funded by EU, under Grant agreement no. 619437. The SUNSEED
project is a joint undertaking of 9 partner institutions and their contributions are
fully acknowledged. The work was also supported in part by the European Research Council (ERC Consolidator Grant no. 648382 WILLOW) within the Horizon 2020 Program.

Also, thanks to Kasper~F.~Trillingsgaard for constructive comments and suggestions.

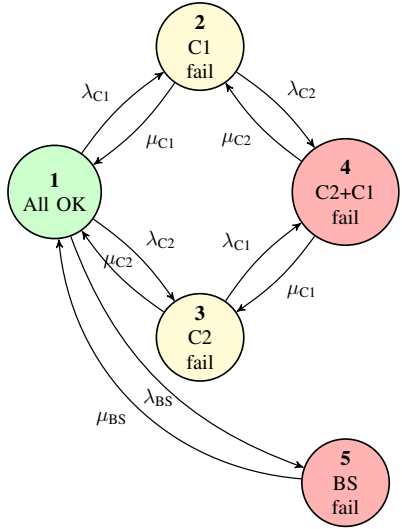
\begin{figure}[htb]
	\centering
	\resizebox{0.6\linewidth}{!}{
\begin{tikzpicture}[->, >=stealth', auto, semithick, node distance=3.5cm]
\tikzstyle{every state}=[fill=white,draw=black,thick,text=black,scale=1,align=center]
\node[state]    (A)[fill=green!20]   {\textbf{1}\\All OK};
\node[state]    (B)[above right of=A,fill=yellow!20]   {\textbf{2}\\C1\\fail};
\node[state]    (C)[below right of=A,fill=yellow!20]   {\textbf{3}\\C2\\fail};
\node[state]    (D)[below right of=B,fill=red!30]   {\textbf{4}\\C2+C1\\fail};
\node[state]    (E)[below right of=C,fill=red!30]   {\textbf{5}\\BS\\fail};
\path
 (A) edge[bend left=10]     node{$\lambda_\text{C1}$}         (B)
 (B) edge[bend left=10]     node{$\mu_\text{C1}$}         (A)
 (A) edge[bend left=10]     node{$\lambda_\text{C2}$}         (C)
 (C) edge[bend left=10, above]     node{$\mu_\text{C2}$}         (A)
 (A) edge[bend right=25, below]     node{$\lambda_\text{BS}$}         (E)
 (E) edge[bend left=40]     node{$\mu_\text{BS}$}         (A)

 (B) edge[bend left=10]     node {$\lambda_\text{C2}$}         (D)
 (D) edge[bend left=10]     node {$\mu_\text{C2}$}         (B)
 (C) edge[bend left=10]     node {$\lambda_\text{C1}$}         (D)
 (D) edge[bend left=10]     node {$\mu_\text{C1}$}         (C)

;
\end{tikzpicture}}
	\caption{State-transition diagram of the continuous time Markov chain that represents the cellular connections C1 and C2 with correlated failures.}
	\label{fig:5-state_diagram}
\end{figure}

\appendix
This appendix explains the approach used to determine the Markov chain failure and restoration rates for the dependent cellular technologies C1 and C2. For this, we consider the CTMC model corresponding to the cellular subsystem of Fig. \ref{fig:network_diagram}. This subsystem is shown in Fig. \ref{fig:5-state_diagram}.

Initially, we specify the known individual availabilities $A_1$ and $A_2$ as well as the known base station availability $A_\text{BS}$, given in Table \ref{tab:f_and_r_rates}. 
Transitions between states are specified by the failure rates denoted by $\lambda$ and restoration rates denoted by $\mu$. Notice that neither failure rates or restoration rates are known for the considered case study. We have therefore made assumptions in the values of the restoration rates as specified in Table \ref{tab:f_and_r_rates}. 

Given the availabilities $A_\text{1}$, $A_\text{2}$, and $A_\text{BS}$, we determine the state probabilities $\pi_i$ of the states in Fig. \ref{fig:5-state_diagram}, by solving the following linear equation system that explains the relations between the steady state probabilities and availability probabilities:
\begin{equation}
 	\begin{bmatrix}
    0 & 1 & 0 & 1 & 1 \\
    0 & 0 & 1 & 1 & 1 \\
    0 & 1 & -1 & 0 & 0 \\
    0 & 0 & 0 & 1 & -1 \\
    0 & 0 & 0 & 0 & 1 \\
    1 & 1 & 1 & 1 & 1
    \end{bmatrix}
	\begin{bmatrix}
	\pi_1\\
	\pi_2\\
	\pi_3\\
	\pi_4\\
	\pi_5
    \end{bmatrix}
 =	\begin{bmatrix}
	1-A_\text{1}\\
	1-A_\text{2}\\
	0\\
	0\\
	1-A_\text{BS}\\
	1
    \end{bmatrix}.\nonumber
\end{equation}

Having obtained the state probabilities $\boldsymbol{\pi}=[\pi_1 \ldots \pi_5]$, we set up the following balance equations that explain the relations between the failure and restoration rates according to Fig. \ref{fig:5-state_diagram}. The assumed mean restoration rates in Table \ref{tab:f_and_r_rates} are given as input and we can then solve the corresponding linear system:
\begin{equation}
 	\begin{bmatrix}
    -\pi_1 & -\pi_1 & -\pi_1 & \pi_2 & \pi_3 & \pi_5 \\
    \pi_1 & -\pi_2 & 0 & -\pi_2 & \pi_4 & 0 \\
    -\pi_3 & \pi_1 & 0 & \pi_4 & -\pi_3 & 0 \\
    \pi_3 & \pi_2 & 0 & -\pi_4 & -\pi_4 & 0 \\
    0 & 0 & \pi_1 & 0 & 0 & -\pi_5 \\
    -1 & 1 & 0 & 0 & 0 & 0 \\
    0 & 0 & 0 & -1 & 1 & 0 \\
    0 & 0 & 0 & 1 & 0 & 0 \\
    0 & 0 & 0 & 0 & 1 & 0 \\
    0 & 0 & 0 & 0 & 0 & 1 
    \end{bmatrix}
	\begin{bmatrix}
	\lambda_\text{C1}\\
	\lambda_\text{C2}\\
	\lambda_\text{BS}\\
	\mu_\text{C1}\\
	\mu_\text{C2}\\
	\mu_\text{BS}\\
    \end{bmatrix}
 =	\begin{bmatrix}
	0 \\
	0 \\
	0 \\
	0 \\
	0 \\
	0 \\
	0 \\
	\mu_\text{C1} \\
	\mu_\text{C2} \\
	\mu_\text{BS}
    \end{bmatrix}. \nonumber
\end{equation}

Thereby we obtain a set of failure rates $\lambda_\text{C1}$, $\lambda_\text{C2}$, and $\lambda_\text{BS}$ that satisfy the constraints of the system in terms of state probabilities, restoration rates, and balance relations between states.

\bibliographystyle{IEEEtran}

\begin{thebibliography}{10}
\providecommand{\url}[1]{#1}
\csname url@samestyle\endcsname
\providecommand{\newblock}{\relax}
\providecommand{\bibinfo}[2]{#2}
\providecommand{\BIBentrySTDinterwordspacing}{\spaceskip=0pt\relax}
\providecommand{\BIBentryALTinterwordstretchfactor}{4}
\providecommand{\BIBentryALTinterwordspacing}{\spaceskip=\fontdimen2\font plus
\BIBentryALTinterwordstretchfactor\fontdimen3\font minus
  \fontdimen4\font\relax}
\providecommand{\BIBforeignlanguage}[2]{{%
\expandafter\ifx\csname l@#1\endcsname\relax
\typeout{** WARNING: IEEEtran.bst: No hyphenation pattern has been}%
\typeout{** loaded for the language `#1'. Using the pattern for}%
\typeout{** the default language instead.}%
\else
\language=\csname l@#1\endcsname
\fi
#2}}
\providecommand{\BIBdecl}{\relax}
\BIBdecl

\bibitem{andrews2014will}
J.~G. Andrews, S.~Buzzi, W.~Choi, S.~V. Hanly, A.~Lozano, A.~C. Soong, and
  J.~C. Zhang, ``What will 5g be?'' \emph{Selected Areas in Communications,
  IEEE Journal on}, vol.~32, no.~6, pp. 1065--1082, 2014.

\bibitem{monserrat2015metis}
J.~F. Monserrat, G.~Mange, V.~Braun, H.~Tullberg, G.~Zimmermann, and
  {\"O}.~Bulakci, ``Metis research advances towards the 5g mobile and wireless
  system definition,'' \emph{EURASIP Journal on Wireless Communications and
  Networking}, vol. 2015, no.~1, pp. 1--16, 2015.

\bibitem{ratasuk2015recent}
R.~Ratasuk, A.~Prasad, Z.~Li, A.~Ghosh, and M.~Uusitalo, ``Recent advancements
  in m2m communications in 4g networks and evolution towards 5g,'' in
  \emph{Intelligence in Next Generation Networks (ICIN), 2015 18th
  International Conference on}.\hskip 1em plus 0.5em minus 0.4em\relax IEEE,
  2015, pp. 52--57.

\bibitem{lahetkangas2014achieving}
E.~L\"ahetkangas, K.~Pajukoski, J.~Vihri\"al\"a, G.~Berardinelli, M.~Lauridsen,
  E.~Tiirola, and P.~Mogensen, ``Achieving low latency and energy consumption
  by 5g tdd mode optimization,'' in \emph{Communications Workshops (ICC), 2014
  IEEE International Conference on}.\hskip 1em plus 0.5em minus 0.4em\relax
  IEEE, 2014, pp. 1--6.

\bibitem{tullberg2014towards}
H.~Tullberg, Z.~Li, A.~Hoglund, P.~Fertl, D.~Gozalvez-Serrano, K.~Pawlak,
  P.~Popovski, G.~Mange, and O.~Bulakci, ``Towards the {METIS 5G} concept:
  First view on horizontal topics concepts,'' in \emph{Networks and
  Communications ({EuCNC}), European Conf. on}.\hskip 1em plus 0.5em minus
  0.4em\relax IEEE, 2014, pp. 1--5.

\bibitem{dahlman20145g}
E.~Dahlman, G.~Mildh, S.~Parkvall, J.~Peisa, J.~Sachs, and Y.~Sel{\'e}n, ``5g
  radio access,'' \emph{Ericsson Review}, vol.~6, pp. 2--7, 2014.

\bibitem{qadir2015exploiting}
J.~Qadir, A.~Ali, K.-L.~A. Yau, A.~Sathiaseelan, and J.~Crowcroft, ``Exploiting
  the power of multiplicity: a holistic survey of network-layer multipath,''
  \emph{arXiv preprint arXiv:1502.02111}, 2015.

\bibitem{apostolopoulos2000reliable}
J.~G. Apostolopoulos, ``Reliable video communication over lossy packet networks
  using multiple state encoding and path diversity,'' in \emph{Photonics West
  2001-Electronic Imaging}.\hskip 1em plus 0.5em minus 0.4em\relax
  International Society for Optics and Photonics, 2000, pp. 392--409.

\bibitem{rausand2004system}
M.~Rausand and A.~H{\o}yland, \emph{System reliability theory: models,
  statistical methods, and applications}.\hskip 1em plus 0.5em minus
  0.4em\relax John Wiley \& Sons, 2004, vol. 396.

\bibitem{strom20155g}
E.~G. Str{\"o}m, P.~Popovski, and J.~Sachs, ``5g ultra-reliable vehicular
  communication,'' \emph{arXiv preprint arXiv:1510.01288}, 2015.

\bibitem{stefanovic2014sunseed}
C.~Stefanovic, P.~Popovski, L.~Jorguseski, and R.~Sernec, ``{SUNSEED}--an
  evolutionary path to smart grid comms over converged telco and energy
  provider networks,'' in \emph{Wireless Communications, Vehicular Technology,
  Information Theory and Aerospace \& Electronic Systems (VITAE), 2014 4th
  International Conference on}.\hskip 1em plus 0.5em minus 0.4em\relax IEEE,
  2014, pp. 1--5.

\bibitem{ross1996stochastic}
S.~M. Ross \emph{et~al.}, \emph{Stochastic processes}.\hskip 1em plus 0.5em
  minus 0.4em\relax John Wiley \& Sons New York, 1996, vol.~2.

\bibitem{borella1997self}
M.~S. Borella, S.~Uludag, G.~B. Brewster, and I.~Sidhu, ``Self-similarity of
  internet packet delay,'' in \emph{Communications, 1997. ICC'97 Montreal,
  Towards the Knowledge Millennium. 1997 IEEE International Conference on},
  vol.~1.\hskip 1em plus 0.5em minus 0.4em\relax IEEE, 1997, pp. 513--517.

\bibitem{sunseed2014web}
\BIBentryALTinterwordspacing
{SUNSEED FP7 project}. [Online]. Available: \url{http://sunseed-fp7.eu/}
\BIBentrySTDinterwordspacing

\end{thebibliography}


\end{document}